\documentstyle[11pt,newpasp,twoside,epsf]{article}
\def\edcomment#1{\iffalse\marginpar{\raggedright\sl#1\/}\else\relax\fi}

\reversemarginpar

\begin{document}

\title{External Shock Model for the Prompt Phase of Gamma Ray Bursts: 
Implications for GRB Source Models}

\author{Charles D.\ Dermer}

\affil{E. O. Hulburt Center for Space Research, Code 7653,
Naval Research Laboratory, Washington, DC 20375-5352}

\author{Kurt E.\ Mitman}

\affil{University of Virginia, Charlottesville, VA, and
Pembroke College, University of Cambridge, Cambridge, England CB2 1RF }

\begin{abstract}
An external shock model for the prompt gamma-ray luminous phase of
gamma-ray bursts (GRBs) is treated both analytically and
numerically. A widely cited derivation claiming that an external shock
model for rapidly variable GRBs must be very inefficient employs an
incorrect expression for the angular timescale. Numerical results show
that variable GRBs can be formed with $> 10$\% efficiency to transform
the directed kinetic energy of the relativistic fireball to
gamma-rays. Successes of the external shock model and difficulties
with an internal wind/colliding shell model are summarized. An
impulsive external shock model is consistent with the supranova model.
\end{abstract}

\section{Introduction}

The prompt gamma-ray luminous phase of GRBs lasts from seconds to
minutes after the start of a burst and provides our best probe
of the processes that take place in the vicinity of the GRB central
engine. Within the framework of the relativistic blast-wave/fireball
scenario, a central problem concerns the nature of the prompt
radiation. In the internal shock model, an active central engine is
assumed to eject waves of relativistic plasma that collide with each
other to form shocks that accelerate nonthermal particles and radiate
high-energy photons. In the impulsive external shock model, a single
relativistic wave of plasma interacts with inhomogeneities in the
surrounding medium to form external shocks that accelerate particles which
radiate the prompt gamma rays.

The resolution of this problem has important implications for theories
of the central engine. In the collapsar model (e.g., MacFadyen,
Woosley, and Heger 2001), the core of a massive rotating star
collapses directly to a black hole, forming a disk that is assumed to
accrete onto the black hole over timescales of seconds to
minutes. This active central engine drives a jet of baryon-dilute
relativistic ejecta that penetrates the surrounding stellar envelope
to form colliding relativistic shells. If the collapsar model is
correct, than an impulsive external shock model is ruled out, because
a single explosive event would be quenched by the massive envelope
surrounding the accreting black hole. No highly relativistic shock
could emerge to form either the prompt or afterglow emission.

By contrast, if an impulsive external shock model is correct, then the
collapsar model is ruled out, because the collapsar model requires an
active central engine and the formation of a jet that persists on the
timescale of the prompt phase.

An impulsive external shock model is, however, compatible with the
supranova model for GRBs (as is an internal shell model). In the
supranova model (Vietri and Stella 1998), a massive star undergoes a
two-step collapse to a black hole. In the first step, a supernova
ejects a remnant, leaving behind a neutron star which is
stabilized against direct collapse to a black hole by its rapid
rotation. The loss of angular momentum support by gravitational and
electromagnetic dipole radiation leads to the collapse of the neutron
star to a black hole some weeks to years after the initial
supernova. The process of black-hole formation drives a baryon-dilute
outflow that interacts with the surrounding supernova remnant material
through an external shock to form GRB radiation.

The purpose of this paper is to show that an impulsive external shock
model can account for the highly variable GRB light curves.

\section{The External Shock Model for GRBs}

Three major criticisms have been directed at an external shock model
for GRBs during the prompt phase. These are:

\begin{enumerate}
\item Highly variable light curves cannot be made with high efficiency;
\item Pulse widths should spread with time; and
\item Gaps in light curves cannot be formed.
\end{enumerate} 

\noindent We answer each criticism below.

\subsection{Short Timescale Variability}

Consider an explosion that produces a spherical shell of relativistic
particles moving with Lorentz factor $\Gamma_0\gg 1$. Fenimore,
Madras, and Nayakshin (1996) showed that if the shell was
instantaneously illuminated over all parts of its surface within the
Doppler beaming cone on angular scales $\theta \la 1/\Gamma_0$, then a
characteristic emission profile is formed with mean duration $t_{FWHM}
\approx 0.2 R/\Gamma_0^2 c$, due to light-travel time delays from
different portions of the surface of the shell. Here $R$ is the
distance of the shell from the explosion center. For a single
relativistic shell moving to larger radii, successive instantaneous
illuminations would form pulses with successively larger durations,
contrary to observations. These authors noted that the conclusion that
a single expanding shell cannot produce a variable GRB could be
avoided if the condition of local spherical symmetry was broken on
size scales $\ll R/\Gamma_0 $, for example, by density
inhomogeneities. This was considered unlikely because of the required
large number of such density enhancements.

Dermer and Mitman (1999; hereafter DM99) confirmed by direct numerical
simulation that if clouds with radii $r \ll R/\Gamma_0 $ existed near
GRBs, short timescale variability (STV) in GRB light curves could
indeed be formed. For $\Gamma_0\approx 300$, comparison with GRB pulse
properties implied that density inhomogeneities located $\approx
10^{16}$ cm from the sites of GRBs with sizes $\approx
10^{12}$-$10^{13}$ cm are required.  The discovery of the evolving
prompt absorption feature in GRB 990705 (Amati et al.\ 2000) provides
unexpected support for this model. The variable absorption can be
explained within the context of either resonance scattering (Lazzati
et al. 2001) or photoelectric absorption (B\"ottcher, Fryer, and
Dermer 2002) models if small dense clouds of sizes $\approx 10^{13}$
cm are found $\approx 10^{17}$ cm from the sites of GRBs, consistent
with the numerical expectations for STV. Explanation of X-ray lines in
GRB 011211 (Reeves et al. 2002) also require small, high density
clouds to account for the features detected with XMM.

Sari and Piran (1997) proposed an analytic argument to demonstrate
that an external shock model could produce STV only if it was very
inefficient. Because this argument has been widely cited as a
refutation of the external shock model, we carefully review it and
point out the error contained within the argument.

Consider a blast wave passing through a spherical density
inhomogeneity (or cloud) with size $r \ll R/\Gamma_0$ that is located
at an angle $\theta$ with respect to the line-of-sight to the
observer. The duration of the received pulse of radiation depends on
the light travel-time delays from different portions of the blast wave
as it interacts with the cloud. Photons which are emitted when the
blast wave passes through the near and far sides of the cloud are
received over a {\it radial} timescale
\begin{equation}
t_r = {2r\over \beta_0\Gamma_0{\cal D} c}\cong {r\over \Gamma_0^2 c}\;,
\label{tr}
\end{equation}
where the Doppler factor ${\cal D} = [\Gamma_0(1-\beta_0 \cos
\theta)]^{-1}$, and $\beta_0=\sqrt{1-\Gamma_0^{-2}}$. The radial
timescale varies by a factor $\approx 2$, depending on whether the
cloud is located on-axis or at an angle $\theta \cong 1/\Gamma_0$.

Photons emitted from points defining the greatest angular extent of
the cloud are received over an {\it angular} timescale
\begin{equation}
t_{ang} \cong {r\theta\over c}\;.
\label{tang}
\end{equation}
Eq.\ 2 can easily be derived from special relativity. Note that if $r
\rightarrow R/\Gamma_0$ and $\theta \rightarrow 1/\Gamma_0$, then
$t_{ang} \rightarrow R/\Gamma_0^2 c$, as expected. When $\theta
\approx 1/\Gamma_0$, $t_{ang} \approx \Gamma_0 t_r\gg t_r$. Except for
those few clouds with $\theta \la 1/\Gamma_0^2$ lying almost exactly
along the line-of-sight to the observer, $t_{ang} \gg t_r$.

Sari and Piran (1997) argue that a highly variable light curve is only
possible in an external shock model if the radiative efficiency is
very low. They define a variability index ${\cal V}$, roughly
corresponding to the number of distinct pulses in a GRB light curve,
given by ${\cal V} = T/\Delta T$, where $T$ is the GRB duration and
$\Delta T$ is a typical pulse width. A highly variable GRB can have
${\cal V}\gg 100$. The efficiency $\eta$ to extract energy from a GRB
blast wave is given by the ratio of the total area $A_c \cong N_c\pi
r^2\approx {\cal V} \pi r^2$ subtended by the $N_c$ clouds within the
Doppler beaming cone $\theta \la 1/\Gamma_0$, to the area $A_{bw} =
\pi R^2/\Gamma_0^2$ of the blast wave within the Doppler beaming
cone. Thus $\eta = {\cal V} \pi r^2/(\pi R^2/\Gamma_0^2)$.

They then claim that $\eta \la 1/4{\cal V} \ll 1$, so that a highly
variable light curve with ${\cal V} >> 1$ must be very inefficient. As
can easily be seen, this expression makes use of the relation ${\cal
V} < (R/\Gamma_0)/2r$, which would follow by assuming that the
characteristic duration of a GRB is $T\approx t_{dur} \approx
R/\Gamma_0^2 c$, and that the variability timescale $\Delta t \approx
 r/\Gamma_0 c$. This last approximation makes use of an
expression for $t_{ang}$ (see eq.\ [2]) that is only correct at
$\theta \approx 1/\Gamma_0$.

As was shown by DM99, the clouds located at angles $\theta \ll
1/\Gamma_0$ with respect to the line-of-sight make a disproportionate
contribution to the variability of GRB light curves precisely because
$t_{ang}$ becomes so small --- and therefore the peak flux of a pulse
becomes so large --- for such clouds. The peak pulse flux
\begin{equation}
\phi_{pk} \propto {{\cal D}^{3+\alpha}\over \max(t_r,t_{ang})}\rightarrow 
{c{\cal D}^{3+\alpha}\over r\theta}\;,
\label{phipk}
\end{equation}
where $\alpha$ is the energy spectral index, and the last expression
holds for clouds with $\theta \ga 1/\Gamma_0^2$. Here we have used a
beaming factor appropriate to isotropic synchrotron radiation in the
comoving frame.  Specifically,
\begin{equation}
{\phi_{pk}(\theta = 1/10\Gamma_0)\over \phi_{pk}
(\theta = 1/\Gamma_0)} \cong 10\cdot 2^{3+\alpha} \cong 80-160\;.
\label{ratiophipk}
\end{equation}
Hence 1\% of clouds, at $\theta \cong 1/10\Gamma_0$, produce 8-16\% of
the fluence in very narrow pulses that are $\approx 100\times$
brighter than clouds at $\theta \approx 1/\Gamma_0$. This produces
highly variable light curves with reasonable ($\ga 10$\%) efficiency.

Fig.\ 1 shows new calculations of GRB light curves in an external
shock scenario. We assume that a GRB explodes with apparent isotropic
energy release of $10^{53}$ ergs and $\Gamma_0 = 300$. Clouds, with a
partial covering factor of 10\%, are assumed to radiate 10\% of their
intercepted energy in the form of a Band-type spectrum (if the clouds
are even more radiative, then the pulses will be even
brighter). Clouds are ``uniformly randomly" distributed between
$10^{16}$ and $10^{17}$ cm, that is, the location of the cloud is
randomly selected throughout the volume of the shell by Monte Carlo
methods, provided that the volume of each cloud does not overlap the
volume of another cloud. The underlying assumption is that no spatial
correlations exist between cloud locations. In Fig.\ 1a, all clouds
have the same radius $r = 10^{13}$ cm, and Gaussian noise is added to
the simulation at a level typical of BATSE GRBs. Fig.\ 1b shows a
simulation where clouds are chosen with equal partial covering factor
per logarithmic interval for clouds with sizes between 10$^{12}$ and
$3\times 10^{13}$ cm.  No noise is added in Fig.\ 1b. As can be seen,
there is no difficulty in making highly variable light curves in an
external shock model, even with a 10\% (or larger) partial covering
factor.

\begin{figure}
\vskip-1.5in
\plottwo{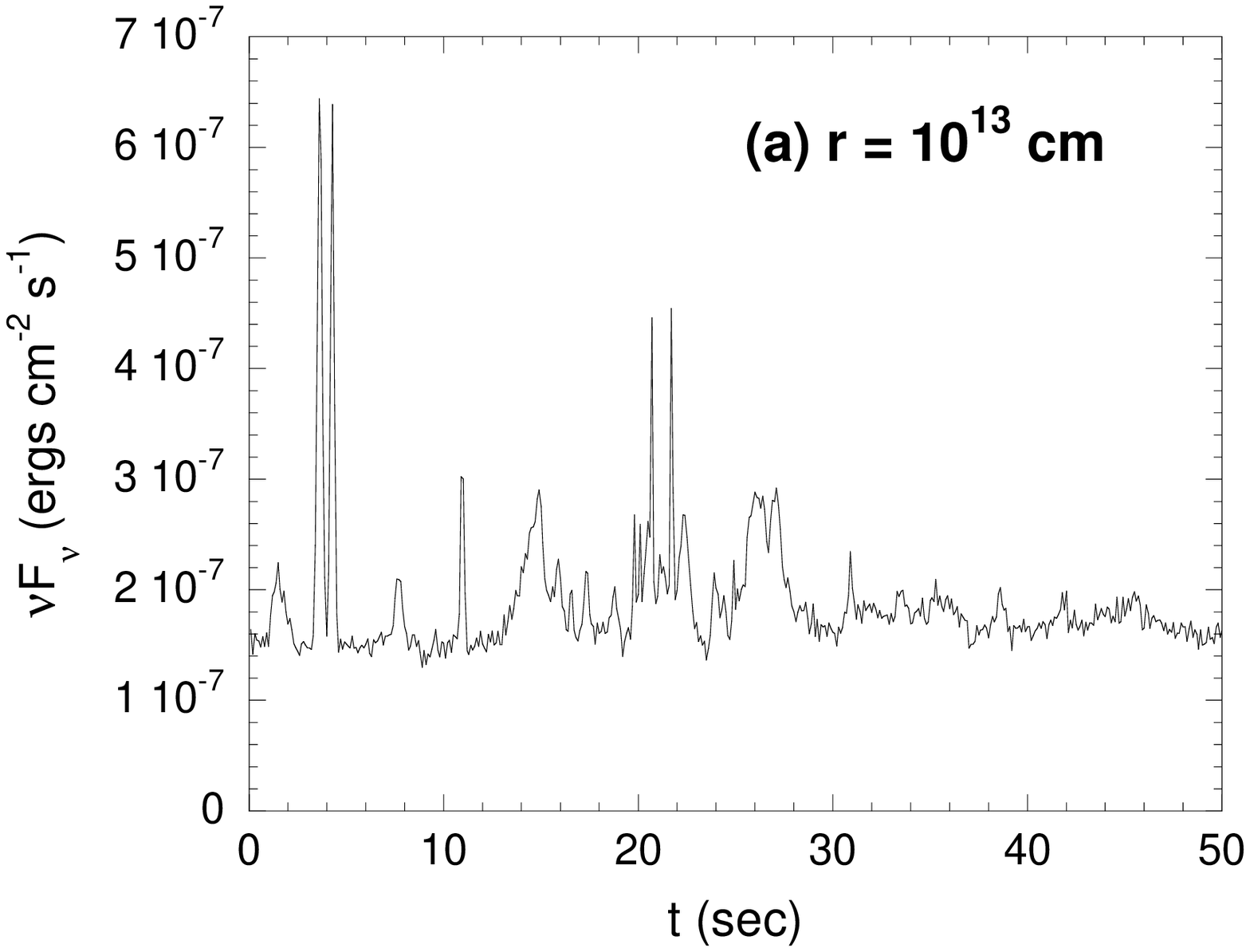}{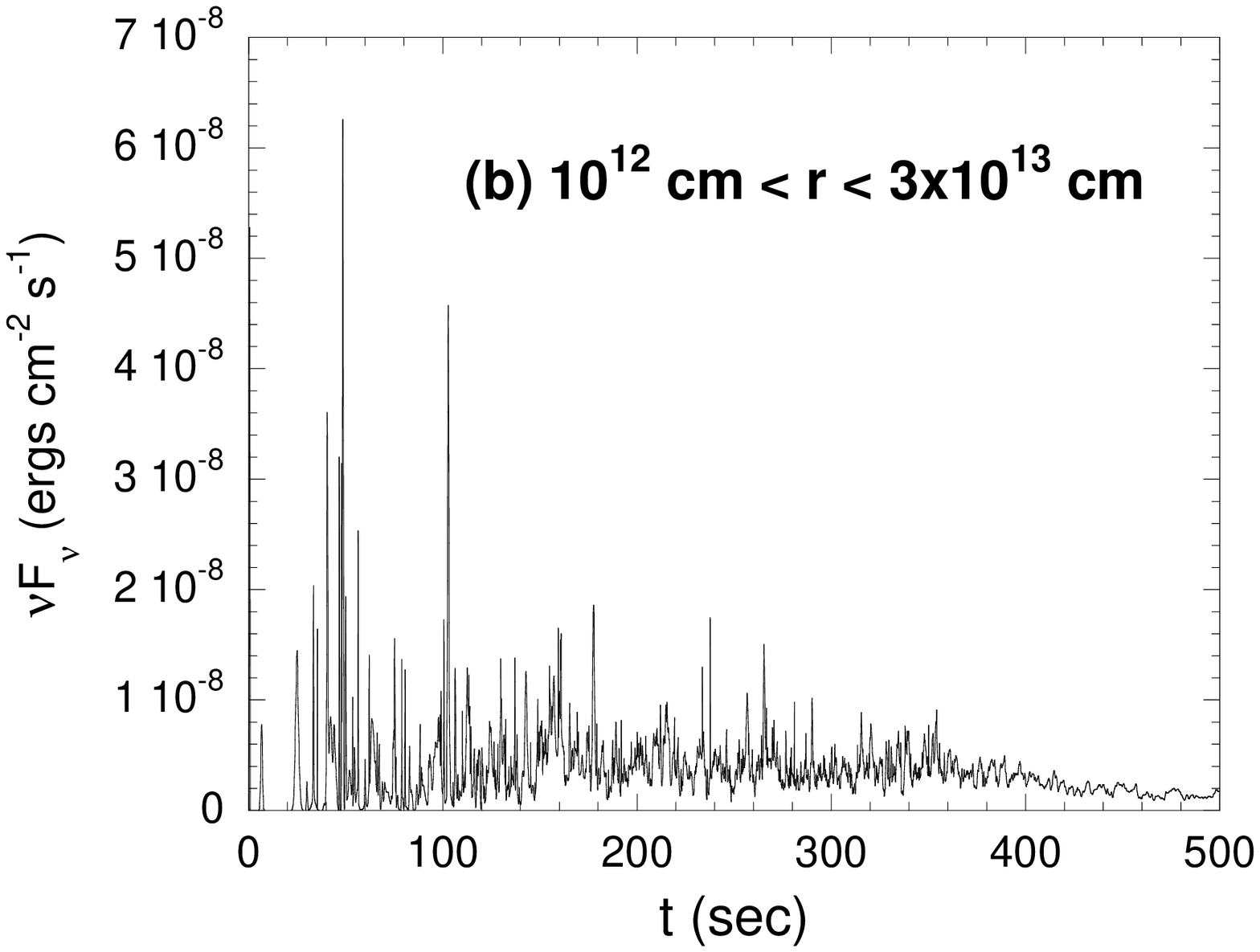}
\caption{Model GRB light curves formed through external shocks with
 clouds in the circumburst medium. (a) All clouds have radii $r =
10^{13}$ cm. (b) Clouds are chosen with equal partial covering factor
per logarithmic interval in cloud size between 10$^{12}$ and $3\times
10^{13}$ cm.}
\end{figure}

\subsection{Pulse Width Spreading}

Another objection to the external shock model is that the pulse widths
should spread with time (Fenimore, Ramirez-Ruiz, and Wu 1999;
Ramirez-Ruiz and Fenimore 2000). One aspect of this criticism consists
of noting that emission from off-axis clouds would arrive later at the
detector, and these would have larger values of $t_{ang}$ than clouds
located along the line-of-sight. This effect is weakly apparent in
Fig.\ 2 of DM99, where a highly idealized scenario is
simulated. No background noise was included, and clouds with identical
radii were uniformly randomly distributed within a shell with discrete
inner and outer boundaries. Figs.\ 1a and 1b separately relax the
first two of these assumptions, with the effect of reducing the
tendency of pulses to spread with time. The more important assumption
of a uniform random distribution has not yet been relaxed. The density
inhomogeneties formed by the interaction of an intense pulsar wind
with a supernova shell would produce Rayleigh-Taylor instabilities and
complex cloud distributions (e.g., Jun 1998) with correlations among
their locations. The uniform random assumption is clearly
oversimplified, and a realistic cloud geometry would further
ameliorate the pulse-width spreading problem (which is in any case
minor, as shown by our simulations). One might furthermore speculate
that the $-5/3$ slope found in the power density spectrum of GRB light
curves (Beloborodov, Stern, and Svensson 1998) is related to the
development of a Kolmogorov spectrum of cloud sizes through
hydrodynamic turbulence.

A second aspect of this objection is that the pulse widths will spread
with time as a result of blast wave deceleration, inasmuch as the
blast wave will progressively slow as it sweeps up material from the
external medium. This is not a problem in a highly structured medium,
however, because whenever the blast wave encounters a cloud with a
sufficiently large column density to produce a bright pulse, that
portion of the blast wave is so strongly decelerated that any further
interactions would produce undetectable emission. The remaining parts
of the blast wave continue to travel with their original speed until
encountering a density inhomogeneity with a ``thick column" that would
produce a bright pulse while only decelerating the intercepted portion
of the blast wave.

\subsection{Gaps in Light Curves}

It has also been argued (Fenimore and Ramirez-Ruiz 1999) that gaps in
GRB light curves are better explained with an active central engine
that falls dormant for some period of time than with an external shock
model. Active regions at different portions of the shell were argued
to have to ``conspire" in an external shock model to form a gap, since
the arrival time to the observer of the different emitting regions on
the shell would depend on the unknown direction to the observer. This
argument does not take into account the strong dependence of pulse
intensity on angle $\theta$ of the cloud, eq.\ (3), which favors those
clouds nearly along the line-of-sight to the observer. Nor does it
consider possible gaps and clustering in cloud distributions. Because
of the strong sensitivity of $\phi_{pk}$ on $\theta$, gaps in light
curves would simply reflect layers of clouds, to reveal a tomographic
image of the circumburster medium, as noted in the conclusion of DM99.

\section{External vs.\ Internal Shocks} 

\subsection{External Shock Model Successes}

Explanations of the prompt GRB emission in terms of an external shock
model offers many advantages over the colliding shell model. First is
a simple understanding of the typical duration timescales of GRBs,
which range from fractions of a second to hundreds of seconds. As
originally pointed out by Rees and M\'esz\'aros (1992, 1993), the
duration of the luminous prompt emission is on the deceleration time
scale $t_d \cong 10 (E_{52}/n_0\Gamma_{300}^8)^{1/3}$ s for explosions
occurring in a uniform circumburster medium with density $n_0$. Here
the apparent isotropic explosion energy is $10^{52}E_{52}$ ergs, and
$\Gamma_{300} = \Gamma_0/300$.  Deceleration in clumpy media will take
place on a timescale $\Delta R/\Gamma_0^2 c \cong 4
R_{16}/\Gamma_{300}^2$ s, where $R_{16} = \Delta R/(10^{16}$ cm) is
the width of density inhomogeneities.
 
Another advantage of an external shock model is that it provides a
simple explanation for the tendency of the $\nu F_\nu$ $E_{pk}$
distribution to appear in a narrow energy range near the peak energy
of the effective area of the detector. This effect is understood
(Dermer, Chiang, and B\"ottcher 1999) by considering both the
triggering properties of GRB detectors and the emission properties of
blast waves with different amounts of baryon loading. Dirty fireballs
with large baryon loading and small $\Gamma_0$ produce extended GRB
emisions with small peak fluxes $\phi_{pk}$, and with $E_{pk}$ values
at low energies. These emissions are unlikely to trigger BATSE as a
result of the smaller peak fluxes and larger backgrounds over the
longer timescales. Clean fireballs produce the bulk of their brief
luminous emission at energies well above the energy range where BATSE
is most sensitive. The result is that BATSE is most sensitive to GRBs
with $E_{pk} $ in the BATSE triggering range.

A model that simultaneously explains the BATSE $t_{50}$ duration
distribution, the $E_{pk}$ distribution, and the peak-flux size
distribution was constructed by B\"ottcher and Dermer (2000). The
resulting model predicts a GRB redshift distribution that can be used
to improve the parameters of the model when compared against the
measured GRB redshift distribution. To avoid fine tuning of
$\Gamma_0$, the external shock model predicts that a class of dirty
fireballs must exist (Dermer, B\"ottcher, and Chiang 1999) with
properties similar to the X-ray rich GRBs discovered with Beppo-SAX
(Heise et al.\ 2001).

Qualitative considerations also suggest a simple explanation for the
GRB variability-luminosity correlation (Fenimore and Ramirez-Ruiz
2000; Reichart et al. 2001). Blast waves which interact with small
dense clouds along the line-of-sight produce intense bright peaks with
total fluence that would, had the GRB occurred in a uniform
circumburster medium, be radiated over a much longer timescale and at
a smaller peak flux level. Consequently, the peak fluxes of highly
variable GRB light curves would reach much larger values than smooth
light curves.

\subsection {Internal Shock Model Difficulties}

When two shells collide to form a spherical radiating surface,
light-travel time effects imply a unique temporal relation between the
shell intensity and $E_{pk}$ that depends only on the (measurable)
spectral indices of the GRB pulse (Soderberg and Fenimore 2001). The
evolution of GRB pulses do not follow the predicted trend, implying
that the simplest version of colliding shell physics is incomplete or
incorrect.

The efficiency of colliding shells is of order $\sim 1$\% (Kumar
2000) unless the contrast between the $\Gamma_0$ factors of the
shells is very large (Beloborodov 2000). If the $\Gamma_0$ contrast is
small, a narrow range of $E_{pk}$ may be possible, but if large,
$E_{pk}$ will range over orders of magnitude, thus reducing the
$\gamma$-ray detection efficiency. In either case, therefore, a
colliding shell model is very inefficient. This makes it difficult to
understand the constant-energy reservoir result of Frail et
al.\ (2001). In contrast, the interactions of a single relativistic
blast wave with stationary material will always have the same relative
Lorentz factor $\Gamma_0$, which can account for the rough similarity
between $E_{pk}$ values in different pulses of a GRB light curve
(e.g., Crider et al. 1999).

In models involving colliding shells, no first principles
understanding for the GRB duration distribution has been proposed, as
the temporal variation in a wind model reflects the period of activity
of the central engine, which is arbitrarily assigned. Consequently, no
explanation for the statistical properties of GRBs, namely the
$E_{pk}$, $t_{dur}$, and $\phi_{pk}$ distributions, has been proposed
within the context of this model.

\section {Implications for Source Models}

If an impulsive external shock model is correct, then, as argued
earlier, the collapsar model is ruled out. As shown by analytic
arguments and numerical simulations, highly variable GRBs can be
formed through interactions of a single blast wave with density
inhomogeneities in the surrounding medium. The existence of such
inhomogeneities may be associated with the material from an earlier
supernova explosion. An impulsive external shock model is consistent
with the supranova model, which provides several advantages
to explain GRB afterglow behavior using the standard blast wave physics
approach (K\"onigl and Granot 2002).

An impulsive external shock model, if correct, implies that GRB
explosions are due to compact objects that collapse directly to denser
configurations, without the intermediate formation of an accretion
disk. This implication is in accord with numerical simulations (e.g.,
Saijo et al.\ 2002) showing that the collapse of a rotating core of a
supermassive star does not leave sufficient material to form an
accretion disk that could power a GRB. Models for the formation of an
impulsive GRB include the pair electromagnetic pulse during black hole
formation (Ruffini et al.\ 2001), or pair production through
neutrino/antineutrino annihilation processes during compact object
coalescence (Janka et al.\ 1999). Extension of neutrino calculations
to the collapse of rotating supramassive neutron stars is important to
determine whether such models provide sufficient energy to explain GRB
observations.

\acknowledgments This work is supported by the Office of Naval Research.

\end{document}